\documentclass[twocolumn,showpacs,preprintnumbers,amsmath,amssymb]{revtex4}
\usepackage[dvipdfm]{graphicx}
\usepackage{dcolumn}
\usepackage{bm}
\usepackage{amssymb}

\begin{document}
\title{Enhanced Coherence of Antinodal Quasiparticles in a Dirty d-wave 
Superconductor}

\author{Katsunori Wakabayashi}
\author{T. M. Rice}
\author{Manfred Sigrist}
\affiliation{Institute f\"ur Theoretische Physik, 
ETH-H\"onggerberg, CH-8093 Z\"urich, Switzerland}
\begin{abstract}
Recent ARPES experiments show a narrow quasiparticle peak 
at the gap edge along the antinodal [1,0]-direction 
for the overdoped cuprate superconductors. 
We show that within weak coupling BCS theory
for a d-wave superconductor the s-wave single-impurity scattering cross section vanishes 
for energies $ \omega = \Delta $ (gap edge). 
This coherence effect occurs through multiple scattering off the 
impurity. For small impurity concentrations the spectral function has a pronounced
increase of the (scattering) lifetime for antinodal quasiparticles but shows 
a very broad peak in the nodal direction, in qualitative agreement with experiment and
 in strong contrast to 
the behavior observed in underdoped cuprates. 
\end{abstract} 


\maketitle

It is well known that non-magnetic impurities
have little influence on 
the thermodynamic properties of the conventional s-wave
superconductors in zero magnetic field \cite{anderson}. 
Non-magnetic impurities are,
however, detrimental to unconventional 
superconductors with higher angular momentum such as d-wave
superconductors. For such superconductors momentum
conservation is essential, due to the anisotropic structure of the
pair wavefunctions. 
%
%
%
In this letter we show that an enhanced coherence appears however
in these superconductors for quasiparticles with an energy at the
gap maximum. In fact such quasiparticles are not scatterered at all 
by a single impurity and are only weakly scattered in the presence of 
a finite density of impurities. This result has similarities to the
cancellation effect found by Zhu et. al.\cite{zhu} in elastic forward scattering
of quasiparticles off a finite density of impurities although they used
a self-consistent Born approximation which does not reproduce the
suppression by multiple scattering off a single impurity.
Because of this coherence effect a quasiparticle peak appears at the
Fermi wavevector in a dirty d-wave superconductor in the antinodal
direction but not at the Fermi wavevector in the nodal direction.
Such behavior was reported recently in angle resolved photoemission
spectra (ARPES) on strongly overdoped Tl-cuprates 
by Plat\'{e} et. al.\cite{plate}
who commented on the striking contrast to underdoped cuprates where
coherence is observed in the nodal direction.
We study the case of point scatterers in a weak coupling d-wave
superconductor. Strong screening and weak coupling superconductivity
are to be expected in these highly metallic overdoped samples.

We discuss briefly the single-impurity
problem in a d-wave superconductor\cite{balatsky,hirschfeld,okuno}. 
The scattering off the impurity is described by the Hamiltonian
\begin{equation}
H_{imp} = \sum_{\bm{k},\bm{k}^\prime,\sigma}V c^\dagger_{\bm{k},\sigma}c_{\bm{k}^\prime,\sigma},
\end{equation}
where $V$ is the coupling strength of the impurity located at
$r=0$,  and 
$c^\dagger_{\bm{k},\sigma}(c_{\bm{k},\sigma})$ is the creation(annihilation)
operator for electron with wave vector $ \bm{k}$ and spin $\sigma$.  
We restrict ourselves here to s-wave scattering.
The effect of the impurity scattering is described by the
T-matrix, $\hat{T}(\omega)$. The Green's function in the presence of an
impurity is 
\begin{equation}
\hat{G}_{\bm{k},\bm{k}^\prime}(\omega) =  \hat{G}^{(0)}_{k}\delta_{\bm{k},\bm{k}^\prime} +
\hat{G}^{(0)}_{\bm{k}}(\omega)\hat{T}(\omega)\hat{G}^{(0)}_{\bm{k}^\prime}(\omega),
\end{equation}
where both $\hat{G}^{(0)}_{\bm{k}}(\omega)$ and $\hat{T}(\omega)$ are
matrices in the Nambu particle-hole space. 
$\hat{G}^{(0)}_{\bm{k}}(\omega)$ is the Green's function in the absence of
impurities, written as 
\begin{equation}
\hat{G}^{(0)}_{\bm{k}}(\omega) =  \left(\omega - \Delta_{\bm{k}}\hat{\rho}_1
- \xi_{\bm{k}} \hat{\rho}_3 \right)^{-1},
\end{equation}
where $\xi_{\bm{k}}$ is the quasi-particle energy,
$\Delta_{\bm{k}}=\Delta\cos(2\phi_{\bm{k}})$ is the gap function with $d_{x^2-y^2}$
symmetry and $\phi_{\bm{k}}$ is the angle of the wave vector $\bm{k}$ with respect
to the $k_x$-axis. The matrices $\hat{\rho}_i$(i=1,2,3) are the Pauli
matrices and
$\hat{\rho}_0$ is unit matrix in the particle-hole space.

The T-matrix for the s-wave scattering is expressed by
$\hat{T} = T_0 \hat{\rho}_0 + T_3 \hat{\rho}_3 $ with 
\begin{equation}
T_0 = \frac{G_0}{c^2-G_0^2}, \qquad
T_3 = \frac{-c}{c^2-G_0^2} .
\end{equation}
The parameter $ c = \cot\delta_0$ introduces the
scattering phase shift $\delta_0$, with $c=0$ in the unitarity limit, 
and $c\gg 1$ in the weak scattering limit.
$G_0(\omega) = 1/(2\pi N_0)\sum_{\bm{k}}{\rm Tr}
\hat{G}_{\bm{k}}^{(0)}(\omega)\hat{\rho}_0$, where 
$N_0$ is the density of states (DOS) per spin at the Fermi energy in the
normal state. Explicitly, we can rewrite $G_0$ as 
\begin{eqnarray}
G_0(\omega)  =
 -\left\langle\frac{\omega}{\sqrt{\Delta_{\bm{k}}^2-\omega^2}}\right\rangle 
= -  \frac{2}{\pi}\frac{\omega}{\sqrt{1-\omega^2}}
 K\left(\frac{1}{1-\omega^2}\right). 
\end{eqnarray}
Here $\langle\cdots\rangle$ means the average over the angle $\phi_{\bm{k}}$,
and $K(x)$ is the
complete elliptic integral of the frist kind \cite{abramowitz}.
Note that the  quasiparticle DOS in the d-wave superconductor
is given by $N(\omega)=-{\rm Im}(G_0(\omega))$. It diverges 
logarithmically at $\omega/\Delta=1$ (see the case of $\Gamma=0$ in Fig.1).
The divergence of $G_0(\omega)$
at the gap edge ( $\omega/\Delta=1$ ) implies that both $T_0$ and
$T_3$ vanish. 
The scattering cross section which is proportional to the square of the
modulus of the $T-$matrix $ T_0 $, vanishes at the quasiparticle
energy $\omega = \Delta$. This remarkable result is a coherence effect
as it only appears with multiple scattering, but not for the Born
limit, i.e. $ c \to \infty $.


We proceed to discuss this interference feature in the presence of
a finite small concentration of impurities \cite{AG,maki}.
After averaging over
all impurity configurations, the translational symmetry in 
the system is effectively restored, and we can rewrite the Green's function,
\begin{eqnarray}
\hat{G}_{\bm{k}}(\omega) 
& =  \left(\omega - \Delta_{\bm{k}}\hat{\rho}_1
- \xi_{\bm{k}}\hat{\rho}_3 -\hat{\Sigma}\right)^{-1},\\
& =  \left(\tilde{\omega} - \tilde{\Delta}_{\bm{k}}\hat{\rho}_1
- \tilde{\xi}_{\bm{k}}\hat{\rho}_3 \right)^{-1}.
\end{eqnarray} 
The self-energy $\hat{\Sigma}$ includes the effects of
impurity scattering, and $\tilde{\omega}$ and $\tilde{\Delta}_{\bm{k}}$
are the renormalized frequency and gap function, respectively.
We decompose the self-energy into
$\hat{\Sigma}=\Sigma_0\hat{\rho}_0+\Sigma_1\hat{\rho}_1+\Sigma_3\hat{\rho}_3$, 
and obtain 
\begin{eqnarray}
\tilde{\omega}=\omega-\Sigma_0, \quad
\tilde{\Delta}_{\bm{k}}=\Delta_{\bm{k}}-\Sigma_1, \quad
\tilde{\xi}_{\bm{k}}=\xi_{\bm{k}}-\Sigma_3.
\end{eqnarray}
Under particle-hole symmetry $\Sigma_3$ vanishes.
In a  conventional s-wave superconductor without magnetic impurities, 
both $\tilde{\omega}$ and $\tilde{\Delta}$ are renormalized in the same
way, resulting in the absence of pair breaking. 
However, in a unconventional
superconductor, there 
is no renormalization in $\Delta$,
i.e. $\tilde{\Delta}_{\bm{k}}=\Delta_{\bm{k}}$, because $ \Sigma_1 $ is zero.  
The self-energy for various cases is summarized in the table
I. and II.
\begin{figure}[h]
\includegraphics[width=\linewidth]{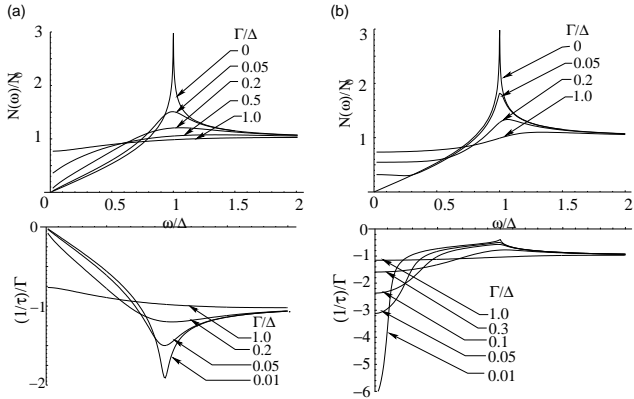} 
\caption{(a) The DOS (upper panel) and
the lifetime(lower panel)  for
 $\Gamma/\Delta$ in the Born limit.
(a) The same plot for the unitarity limit.}
\end{figure}

Next we compare the two scattering limits, the Born and the unitarity
limit. The former yields the renormalized frequency
$\tilde{\omega}$, given by \cite{puchkaryov}
\begin{equation}
\tilde{\omega} = \omega + \Gamma 
\left\langle
\frac{\tilde{\omega}}{\sqrt{\Delta^2_{\bm{k}}-\tilde{\omega}^2}}
\right\rangle.\label{eq:romega}
\end{equation}
Here $\Delta_{\bm{k}}=\Delta\cos(2\phi_{\bm{k}})$, $\Gamma=n_i/\pi N_0$
 and  $n_i$ is the 
impurity concentration. Note that $ \Gamma^{-1} $ corresponds to the
quasiparticle lifetime close to the Fermi energy in the normal metal
state.
The DOS  is obtained by
\begin{eqnarray}
\frac{N(\omega)}{N_0} 
= {\rm Im} \left\langle
            \frac{\tilde{\omega}}{\sqrt{\Delta^2_{\bm{k}}-\tilde{\omega}^2}}
           \right\rangle, 
\end{eqnarray}
where $\tilde{\omega}$ is self-consistently determined
by Eq.(\ref{eq:romega}) for given 
$\omega/\Delta$ and $\Gamma/\Delta$.
In the upper panel of Fig.1 (a), 
the DOS is shown for several values of $\Gamma$. 
For pure samples ($\Gamma=0$) the DOS logarithmically diverges 
at the gap edge ($\omega/\Delta=1$). This singularity is removed, 
however, by the presence of potential scattering, and the progressively 
weaker maximum moves
toward lower energy with increasing $ \Gamma $. 

The lifetime of the quasi-particle, $\tau$, is deduced from the imaginary part
of the self-energy
\begin{eqnarray}
1/\tau={\rm Im}[\tilde{\omega}_p]={\rm Im}[\Sigma_0].
\end{eqnarray}
at the pole
$\tilde{\omega}_p$ of the Green's 
function given by the zero of the denominator:
\begin{eqnarray}
\tilde{\omega}^2_p - \Delta^2_{\bm{k}} - \xi_{\bm{k}}^2 = 0.
\end{eqnarray}
In the lower panel of the Fig.1(a), 
the inverse lifetime as a function of $\omega$ is depicted 
in the Born limit. Note that the $1/\tau\propto -N(\omega)$ and the lifetime
increases towards lower frequencies $\omega$ in this limit.
Towards $\omega\rightarrow\Delta$ and $n_i\rightarrow 0$ the
lifetime vanishes due to the divergence in the DOS.

\begin{table}
\caption{\label{tab:table1}
The self-energy term for the Born limit. 
``Imp.'' indicates the type of impurities, 
i.e. N(M) is non-magnetic (magnetic) impurities.
The min($\tau^{-1}$) means the value of $\omega$ which gives
minimum value of the inverse lifetime $\tau^{-1}$.
Here $1/\tau_1$ and $1/\tau_2$ are the parameter which represents 
pair-breaking caused by non-magnetic and magnetic impurity,
respectively. $u=\tilde{\omega}/\tilde{\Delta}$.
}
\begin{ruledtabular}
\begin{tabular}{ccccc}
 &  Imp. & $\Sigma_0$ & $\Sigma_1$ & min($\tau^{-1}$) \\
d-wave & N& $\Gamma \left\langle
\frac{\tilde{\omega}}{\sqrt{\Delta^2_{\bm{k}}-\tilde{\omega}^2}}
\right\rangle$ &0&0 \\
s-wave$^1$ & N &
     $-\frac{1}{2\tau_1}\frac{u}{\sqrt{1-u^2}}$
 &$+\frac{1}{2\tau_1}\frac{1}{\sqrt{1-u^2}}$& $\Delta$ \\
s-wave$^2$ & M &
     $-\frac{1}{2\tau_2}\frac{u}{\sqrt{1-u^2}}$
 &$-\frac{1}{2\tau_2}\frac{1}{\sqrt{1-u^2}}$&0 \\
\end{tabular}
\end{ruledtabular}
\end{table}

\begin{table*}
\caption{\label{tab:table2}
The self-energy term for the unitarity limit. 
In this table, ``Imp.'' indicates the type of impurities, 
i.e. N(M) is non-magnetic (magnetic) impurities.
The min($\tau^{-1}$) means the value of $\omega$ which gives
minimum value of the inverse lifetime $\tau^{-1}$.
Here $\tau_s=\tau_2(1+\zeta^2)^2$, $\gamma=|(1-\zeta^2)/(1+\zeta^2)|$,
$u=\tilde{\omega}/\tilde{\Delta}$ and $\zeta=(J/2)S\pi N_0$.
}
\begin{ruledtabular}
\begin{tabular}{ccccc}
& Imp. & $\Sigma_0$ & $\Sigma_1$ & $min(\tau^{-1}$) \\
d-wave & N & $\Gamma \left\langle
\frac{\tilde{\omega}}{\sqrt{\Delta^2_{\bm{k}}-\tilde{\omega}^2}}
\right\rangle^{-1}$ &0&$\Delta$ \\
s-wave$^1$ & N &
     $-\frac{1}{2\tau_1}\frac{u}{\sqrt{1-u^2}}$
 &$\frac{1}{2\tau_1}\frac{1}{\sqrt{1-u^2}}$&$\Delta$ \\
s-wave$^2$ & M &
     $-\frac{1}{\tau_s}\frac{u}{\sqrt{1-u^2}}\frac{u^2-1}{u^2-\gamma^2}(1+\zeta^2)$ 
 &$-\frac{1}{\tau_s}\frac{u}{\sqrt{1-u^2}}\frac{u^2-1}{u^2-\gamma^2}(1-\zeta^2)$
&$\Delta$ \\
\end{tabular}
\end{ruledtabular}
\end{table*}

Now we turn to  the unitarity limit\cite{pethick,hotta,sun} where
the renormalized 
frequency is obtained as
\begin{equation}
\tilde{\omega} = \omega + \Gamma 
\left\langle
\frac{\tilde{\omega}}{\sqrt{\Delta^2_{\bm{k}}-\tilde{\omega}^2}}
\right\rangle^{-1}.
\label{eq:romega2}
\end{equation}
The DOS in the unitarity limit is determined through
the Eq.(\ref{eq:romega2}) for given
$\omega/\Delta$ and $\Gamma/\Delta$. Similar to the Born limit
impurity scattering removes the logarithmic singularity in the DOS of
the pure system (Fig.1 (b)). In contrast to the Born limit the maximum
moves toward higher energies with increasing $\Gamma$. 
\begin{figure*}
\includegraphics[width=\linewidth]{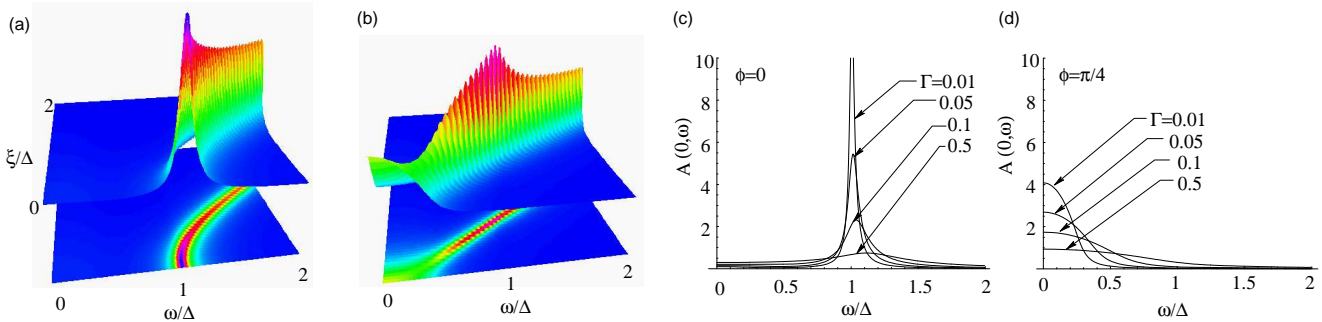} 
\caption{The variation of spectral density $A( \xi_{\bm{k}},\omega)$
for $0<\xi/\Delta<2$ and $\Gamma/\Delta=0.05$ in the unitarity limit,
(a) antinodal ($\phi_{\bm{k}}=0$) (b) nodal direction($\phi_{\bm{k}}=\pi/4$).
A cut of the spectral density along $\xi_{\bm{k}}=0$ for 
$0.05<\Gamma/\Delta <0.5$, (c) antinodal (d) nodal direction.
}
\end{figure*}

In the lower panel of the Fig.1(b), the lifetime as a function
of $\omega$ is shown for the unitarity limit, where $1/\tau\propto
-1/N(\omega)$. As anticipated from the single-impurity discussion the
the quasiparticle lifetime increases at $\omega/\Delta=1$ compared to
the normal state value. While in the single-impurity case at
$\omega = \Delta $ scattering is completely absent, the effect on the
quasiparticle life time becomes less pronounced with increasing
impurity concentration. The reason lies in the broadening of the
quasiparticle spectrum, such that the condition of $ \omega =
\Delta $ cannot be perfectly 
satisfied by a quasiparticle state. Clearly this effect is most significant in
the low impurity concentration limit where the spectral width of the quasiparticle narrows. 
It is also interesting to see that the lifetime behaves very differently
for the two scattering limits for energies much lower than $\Delta$. 
The dramatic decrease of the lifetime 
towards $ \omega=0 $ in the unitarity limit 
is connected with the presence of a zero-energy
boundstate at the impurity which gives rise to resonant scattering. 
Again the broadening of the quasiparticle spectrum with increasing
impurity concentration makes this feature less pronounced. In the Born
limit obviously this feature is absent, as the Born limit does not
capture the boundstate which is a result of multiple scattering at the
impurity. In this case actually the opposite happens, the lifetime
increases as the quasiparticle energy goes towards zero.

A possibility to observe this enhanced coherence
lies in ARPES measurements of the quasiparticle spectrum.
We consider the spectral function $A(\xi_{\bm{k}}, \omega)$ 
given by
\begin{equation}
A( \xi_{\bm{k}}, \omega) = -\frac{1}{\pi} {\rm Im} G_{11}(\xi_{\bm{k}}, \omega).
\end{equation}
There are two distinct directions, nodal and anti-nodal
momentum directions. In Fig.2, the spectral densities of the (a)
anti-nodal ($\phi_{\bm{k}}=0$) 
and (b) nodal ($\phi_{\bm{k}}=\pi/4$) direction are shown
for $\Gamma/\Delta=0.1$.
The longer lifetime gives rise to sharper peaks in the spectral
density, which we see as peak narrowing and rising around $\omega/\Delta=1$.
for both directions. In addition for the nodal direction the lower energy
states become considerably broadened. 
In Fig.2 (c) and (d), the dependence of the spectral density
on the pair breaking parameter at the Fermi wavevector ($k_F$), i.e.
$\xi_{\bm{k}}=0$ is shown.
The increase of the impurity concentration strongly suppresses 
these peak structure as shown.


Finally we would like to interpolate between the Born and unitarity 
limits by a varying coupling strength\cite{mineev}.
The selfenergy $\Sigma_0$ is given by 
\begin{equation}
\Sigma_0 =
 \Gamma\frac{G_0(\tilde{\omega})}{\cos^2\delta_0
-\sin^2\delta_0G^2_0(\tilde{\omega})}, 
\end{equation}
where  $\delta_0$ is the (s-wave) scattering phase shift.
In Fig.3 (a) and (b), the phase shift dependence of lifetime for 
various values of $\Gamma$ is shown at $\omega/\Delta=0$ and $\omega/\Delta=1$,
respectively.
At $\omega/\Delta=1$, the enhancement of the lifetime grows
with increasing the phase shift, in the low impurity 
concentration regime. On the other hand, at $\omega/\Delta=0$,
the lifetime rapidly shrinks with increasing phase shift. This is the
effect of the bound state whose energy is determined by the condition
$ G_0(\omega) = \cot \delta_0 $ and which approaches zero for $ \delta_0 \to
\pi/2 $. 
\begin{figure}[h]
\includegraphics[width=\linewidth]{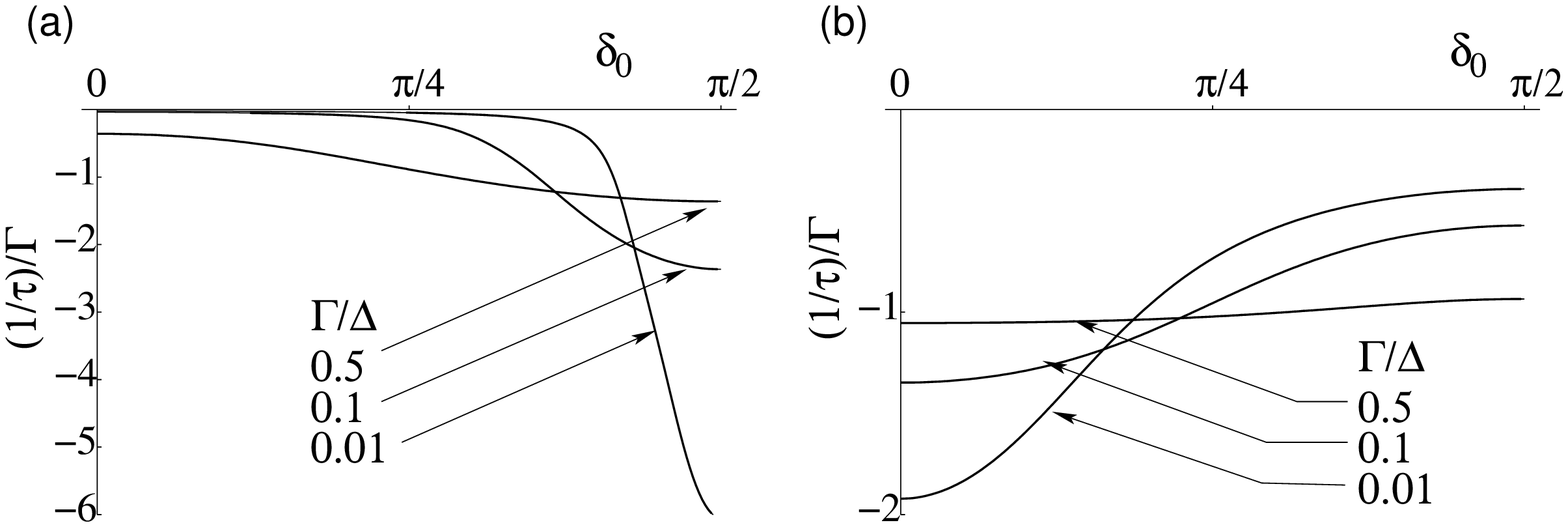} 
\caption{The phase shift dependence of 
the lifetime at (a) $\omega/\Delta=0$ and (b) $\omega/\Delta=1$.}
\end{figure}

Finally, we shall discuss the s-wave superconductor with 
non-magnetic or magnetic impurities in both Born and unitarity
limit\cite{shiba}. 
In order to discuss the magnetic impurities we introduce the
4$\times$4 Green's function including spin degrees of
freedom\cite{maki,shiba}. 
The magnetic impurities in the superconductor are classical, and
interact with electrons through the s-d Hamiltonian
\begin{eqnarray}
H_{sd}=-\frac{J}{2N}\sum_{\bm{k},\bm{k}^\prime}C_{\bm{k}}^\dagger
{\boldsymbol \alpha}C_{\bm{k}^\prime}\cdot {\bf S},
\end{eqnarray}
where $C_{\bm{k}}=(c_{\bm{k}  \uparrow}^\dagger,c_{\bm{k} \downarrow}^\dagger,
c_{-\bm{k}\uparrow},c_{-\bm{k}\downarrow})$, and
the electron spin operator
\begin{eqnarray}
{\boldsymbol \alpha}=\frac{1+\rho_3}{2}{\boldsymbol \sigma}+\frac{1-\rho_3}{2}
\sigma_2{\boldsymbol \sigma}\sigma_2,
\end{eqnarray}
with $\hat{\sigma}_i$(i=1,2,3) as the Pauli
matrices in spin space.
Then we obtain the renormalized frequency and gap function as
\begin{eqnarray}
\tilde{\omega}=\omega-\Sigma_0, \quad
\tilde{\Delta}_{\bm{k}}=\Delta_{\bm{k}}-\Sigma_1.
\end{eqnarray}
The expression of $\Sigma_0$ and  $\Sigma_1$ are shown in table I, II
for both Born and unitarity limits.
Since both frequency and order parameter are renormalized,
the lifetime is given as
\begin{eqnarray}
1/\tau={\rm Im}[\tilde{\omega}_p]
= {\rm Im}[\Sigma_0\pm\sqrt{\xi^2+(\Delta+\Sigma_1)^2}].
\end{eqnarray}
The values of $\omega$ which gives the enhancement of the quasiparticle
lifetime are summarized in tables I and II.
In the unitarity limit, again the  enhancement of the quasiparticle lifetime is observed  at the gap edge, i.e. $\omega/\Delta=1$.
On the other hand, in the Born limit, the analogous effect occurs at $\omega/\Delta=0$.
In the case of the s-wave superconductor with non-magnetic
impurities, the frequency and gap function renormalize in
the same way irrespective to the phase shift ($\omega/\Delta=\tilde{\omega}/\tilde{\Delta}$). 
In zero field thermodynamic properties remain unchanged by non-magnetic 
impurities according to Anderson's
theorem\cite{anderson}.

In conclusion, we have investigated the lifetime of the quasiparticles in
d-wave superconductor by using the Green's function techniques with
$T-$matrix approximation. 
A quasi-particle with energy 
$\omega/\Delta=1$ does not suffer scattering from a
single non-magnetic impurity. 
This peculiar feature, which is the manifestation of the 
superconducting coherence of the quasiparticle,
leads to a long lifetime of the quasiparticle
in the regime of diluted density of impurities and
a spectral function which shows a peak narrowing 
in the antinodal direction. 
The best case to look for this effect in the cuprate superconductors
is in the strongly overdoped region. These are the most metallic
samples so the condition of s-wave scattering is best fulfilled.
Simultaneously the superconductivity approaches a weak 
coupling d-wave limit. Very recently Plat\'{e} et.al.\cite{plate} reported an
ARPES study of an overdoped Tl-cuprate. Their Fig.4(b) shows a 
quasiparticle peak in the antinodal but not in the nodal directions in 
contrast to the observations in the underdoped samples but in agreement with the analysis presented here. 

Financial support from Swiss National Science Foundation and  NCCR MaNEP are gratefully
acknowledged.

\end{document}